\documentclass[superscriptaddress, aps, prl, reprint, floatfix] {revtex4-2}

\usepackage{graphicx}
\usepackage{dcolumn}
\usepackage{bm}
\usepackage{amssymb}
\usepackage[dvipsnames]{xcolor}

\newcommand*{\rom}[1]{\expandafter\@slowromancap\romannumeral #1@}

\begin{document}

\title{Slit-snapping and multistability in buckled  beams with partial cuts}
\author{Bernat Durà Faulí}
\author{Lennard Kwakernaak}
\author{Martin van Hecke}
\affiliation{Huygens-Kamerlingh Onnes Laboratory, Universiteit Leiden, PO Box 9504, 2300 RA Leiden, Netherlands}
\affiliation{AMOLF, Science Park 104, 1098 XG Amsterdam, Netherlands}

\date{\today}

\begin{abstract}
Elastic instabilities such as buckling and snapping have
evolved into a powerful design principle, enabling memory, sequential shape morphing, and computing in metamaterials and devices. Modifying the post-buckling configurations or their snapping transitions would greatly expand design possibilities, yet general principles for controlling elastic instabilities are lacking.
Here, we show that adding a partial cut, or slit, to a flexible beam enables precise control of post-buckling behavior: under compression, slit-beams first buckle, then snap, leading to tristability within the hysteretic regime. 
A truss model explains these phenomena by uncovering the interplay of geometric and slit-induced nonlinearities.
Leveraging these insights, we realize multi-slit beams with programmable behavior, unlocking a vast design space featuring giant hysteresis, quadstability, multi-step snapping, tristability at zero compression, and compression-induced snapping between left- and right-buckled states. Our strategy is general, simple to design and implement, and enables mechanical metamaterials and devices with advanced memory and sequential behavior.
\end{abstract}

\maketitle

\subsection*{{INTRODUCTION}}


The main elastic instabilities are buckling, where an initially straight structure spontaneously curves under compression, and snapping, where a curved structure hysteretically changes shape under transverse loading \cite{bertolidmetareview2017}.
Underlying these instabilities are geometric nonlinearities, where large deformations of slender structures dominate
their mechanical response \cite{KatiaReis2010NegativePoisson, Corentin2015DiscontinuousBuckling}. 
While instabilities are important
to avoid in many industrial structures, they are explored in Nature
to power, e.g., the snapping of the Venus flytrap \cite{sachseSnapping2020,holmesSnapping2007} or the jumping of locust and grasshoppers \cite{bennet-clarkEnergetics1975,queathemOntogeny1991}.
This has inspired the use of buckling and snapping
for applications such as rapid actuation, energy damping and harvesting, and switchable materials in soft robots and metamaterials\cite{jinUltrafast2023,overveldeAmplifying2015,yangPhasetransforming2016,yangMultistable2019,dingSequential2022,rafsanjaniBucklinginduced2017,rafsanjaniSnapping2015,shanMultistable2015}.
In particular, the bistability of beams in the post-buckling regime allows them to function as `material bits', which can be explored for mechanical memory \cite{Reis2021MagnetoMechanical}, sequential shape morphing 
\cite{Katia2022,meeussen}
and computing \cite{LennardCounter2023,Jerry2024}.

Constraints on the instabilities,
or the conditions under which they occur, thus limit the range of functionalities. For example, 
a rectangular beam under axial compression buckles but does not snap under increased compression; post-buckling, it takes on only two configurations; and multistability requires compression. While
general principles to modify instabilities are lacking, in particular for altering the number of post-buckling branches or designing the snapping transitions between them, 
we note that contact interactions are another source of mechanical nonlinearities \cite{LennardBumping2024, LennardCounter2023,HolmesOyster2023}.  
This poses the question if and how the repertoire of elastic instabilities can be augmented by combing contact and geometric instabilities within a single slender structure.

\begin{figure}
\centering
\includegraphics{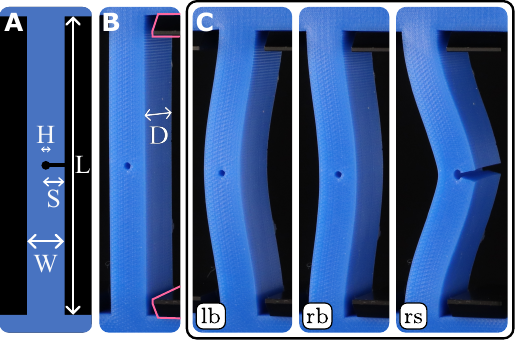}
\caption{\textbf{Slit beams: buckling, snapping and  tristability.} (\textbf{A-B}) Geometry of a slitted beam in its neutral configuration at zero compression ($t=0.125$, $s=0.6$). The beam is terminated with anchoring blocks (blue) and fixed by clamps (black, outline highlighted).
\textbf{(C)} When the compressive strain $\varepsilon$ is increased above the buckling strain
$\varepsilon_b$, 
the beam buckles and takes either a left-buckled (`lb') or right-buckled (`rb')  configuration, both with closed slit. When $\varepsilon$ is increased further, the right-buckled beam beam slit-snaps at $\varepsilon_o$, and this right snapped (`rs') configuration initially remains stable when the strain is decreased, until it snaps close at $\varepsilon_c$. The three panels evidence tristability in the hysteretic range $\varepsilon_c < \varepsilon   < \varepsilon_o$. 
}
\label{fig:beamsketch}
\end{figure}

Here we show that endowing flexible structures with
partial cuts, or {\em slits}, significantly expands their range of mechanical instabilities.
Such slits straightforwardly introduce (contact) nonlinearities: while
compressive stresses keep the slit closed
and preserve the instabilities of the uncut structure, local tensile stresses can open the slit, effectively altering the geometry and enabling additional instabilities. 
Indeed, while slits preserve the buckling instability associated with the closed structure,
the curvature of buckled beams can induce tensile stresses that open the slits, making them ideally 
to control and modify the post-buckling behavior. 

To illustrate this strategy, consider a straight beam with rectangular cross-section, modified with a transverse cut extending from the right edge to its midpoint (Fig.~1). When the compressive strain $\varepsilon$ is increased, this slitted beam initially exhibits buckling and the slit remains fully closed --- inheriting the instability of the closed configuration which corresponds to an ordinary beam
(Fig.~1C). Under increased compression,
a right-buckled beam eventually exhibits a sudden {\em slit-snapping} instability that produces a strongly curved, open configuration (Fig.~1C). This instability is triggered by tensile stresses across the slit produced by the 
increasing curvature of post-buckling beams, and leads to the 
switching from a closed to an open configuration under increased axial compression. 
As we detail below, the transition between open and closed configurations is hysteretic, and leads to a tristable regime where the beams can be buckled to the left, buckled to the right, or 
right-snapped (Fig.~1C).
Hence, slits allows beams to both buckle and snap under compression, and expand their multistability. 

Here we systematically explore the instabilities of slit beams, combining experiments, simulations and theory. First, we
demonstrate how the thickness of the beam and size and position of the slit can be used to precisely control the opening and closing transitions.
We capture slit-snapping and its parameter dependencies in an effective truss model that naturally
combines the bifurcation scenarios for closed and opened beams with the contact-nonlinearities of the slit. We
then use our insights obtained for single-slit beams to 
explore and rationally design beams with multiple slits and more exotic instabilities, where changes in the curvature induced by the opening or closing of one slit lead to interactions between the instabilities of the slits.
For beams with dual slits, we determine the range of slit-snapping scenarios as function of slit location, and experimentally realize beams with giant hysteresis, quadstability, and two-step snapping. Finally, to further show the power of our strategy, we realize two multiple-slit beams with targeted extreme behavior: one that is tristable at zero compression, and one that snaps between left and right buckling branches under compression. 
Our simple, geometric, and general strategy is poised to have a broad impact on the control and optimization of elastic instabilities, opening up exciting possibilities for applications in soft robotics, sensors, metamaterials, and in-materia mechanical computing \cite{KatiaRobot2024, LennardCounter2023, Colin2025pathdependency, Yasuda2021MechanicalComputing}.

\subsection*{{RESULTS}}

\begin{figure}
\includegraphics{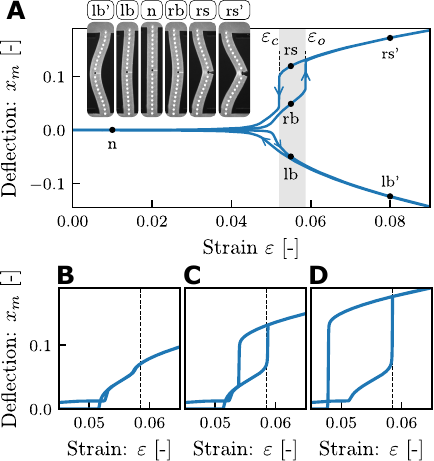}
\centering
\caption{\textbf{Bifurcation diagram and effect of slit size.} 
\textbf{(A)} Bifurcation diagram observed in experiments, with configurations as indicated ($t=0.125$, $s=0.6$, $H=2$ mm). 
(\textbf{{B-D}})  Numerically obtained bifurcation diagrams for $t=0.125$ and (\textbf{B}) $s=0.2$, (\textbf{C}) $s=0.6$ and (\textbf{D}) $s=0.8$. The opening strain $\varepsilon_o=0.0585$ is independent of $s$ (dashed), while the closing strain varies strongly with $s$.
}
\label{fig:slit_trend}
\end{figure}

\subsubsection*{Experiments}
We consider rectangular beams (length $L=80$ mm, depth $D=25$ mm and variable width $W$), cast from Smooth-On Mold Star\texttrademark 30 VPS rubber in 3D printed open-face molds (Fig.~\ref{fig:beamsketch}A,B).
A transversal slit (length $S$) is cut after de-molding using a 
custom made guillotine, and is terminated with a small hole (diameter $H=2$ mm) that prevents tearing. We characterize slitted beams by the dimensionless parameters $t=W/L$ and $s=S/W$. 
The beams are mounted in a custom made compression device that controls the uniaxial compressive strain $\varepsilon:=u_y/L $ by a stepper motor (strain rate 2 mm/min), and records the configuration and dimensionless lateral midpoint deflection $x_m = x/L$ by tracking white painted dots along the centerline of the beams (see Supplemental Information). 

\subsubsection*{Bifurcation diagram}
We now consider the states and transitions for a slitted beam ($t=0.125$, $s=0.6$, $H=2$ mm) by slowly sweeping the compressive strain $\varepsilon$ and characterizing the beam by the mid-beam deflection $x_m$. This yields a bifurcation diagram, the demonstrates
both the  tristability and snapping of slit beams (Fig.~\ref{fig:slit_trend}A and movies S1 and S2).
At small strains, we observe ordinary buckling at $\varepsilon\approx 0.046$. Right after this pitchfork bifurcation, the 'lb' and 'rb' configurations are mirror-symmetric and the slit remains closed (Fig.~\ref{fig:slit_trend}A). Further compression of the 'lb' configuration yields a smooth and reversible increase of $|x_m|$. However, further compressing the 'rb' configuration leads to a sharp snapping transition
at the critical strain $\varepsilon = \varepsilon_o \approx 0.057$, which results in a sharp increase of $x_m$ and produces an open 'rs' configuration. This transition is hysteretic: lowering the strain yields an unsnapping transition from the open to the closed 'rb' configuration at $\varepsilon=\varepsilon_{c} \approx 0.05$. We refer to these phenomena as {\em slit-snapping}. 
In the regime $\varepsilon_c<\varepsilon<\varepsilon_o$, 
the beam is tristable (Fig.~1C,2A).
Hence the slit produces a new solution branch of open configurations, which hysteretically connects to one of the  ordinary buckling branches.

\subsubsection*{Design parameters}
We investigate the dependence of 
the bifurcation diagram and critical strains on the beam aspect ratio $t$ and slit size $s$ by numerical simulations and 
by experiments (see Supplemental Information). 
As expected, the aspect ratio $t$ sets the overall scale of the critical strains, which approximately scale as $t^2$, while the buckling strain $\varepsilon_b$ remains independent of slit size.
Considering the effect of slit size, we observe that a threshold of $s \approx 0.4$ is necessary to initiate the opening transition. 
However, once snapping occurs, $s$ no longer influences the opening strain $\varepsilon_o$.
In contrast, the slit size strongly impacts the closing strain $\varepsilon_c$ and thus controls the size of the hysteresis loop. 
This leads to three different slit-snapping scenarios: for very small $s$, the beam opens smoothly, developing in the the $x_m(\varepsilon)$ curve a kink near $\varepsilon_o$  (Fig.~\ref{fig:slit_trend}B); for intermediate $s$, we obtain the combined buckling-snapping  bifurcation diagram with $\varepsilon_c>\varepsilon_b$ (Fig.~\ref{fig:slit_trend}C). For large $s$, the snapped configuration remains stable below the buckling thresholds, so that  $\varepsilon_c < \varepsilon_b$ 
(Fig.~\ref{fig:slit_trend}D). 
The parameters $t$ and $s$ enable the design of beams with tunable snapping transitions $\varepsilon_o$ and $\varepsilon_c$. The aspect ratio $t$ sets the scale of both $\varepsilon_b$ and $\varepsilon_o$, while $s$ controls $\varepsilon_c$ and the hysteresis width, which approaches a minimum of $\varepsilon_c \approx 0.5 \varepsilon_o$ as $s \rightarrow 1$.

\subsubsection*{Mechanism of slit-snapping}

\begin{figure*}
\centering
\includegraphics[width = \textwidth]{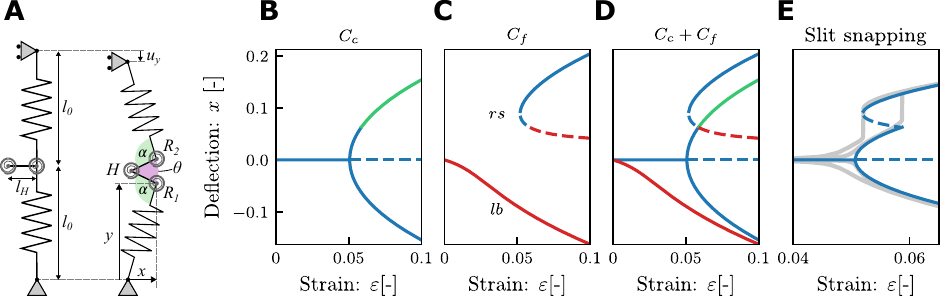}
\caption{\textbf{Slit-truss model.} (\textbf{A}) Geometry at rest (left) and in the open state (right). The geometry and mechanics are specified by the truss lengths $l_0$ and stiffnesses $k_s$, lengths of the bars $l_H$, and parameters of the 
center (left) hinge $H$ (torsional stiffness $\kappa_\theta$, opening angle $\theta$, rest angle zero) and right hinges $R_1$ and $R_2$ (torsional stiffness $\kappa_\alpha$, opening angle $\alpha$, rest angle $\pi/2$).
\textbf{(B-C)} Stable (full) and unstable (dashed) solutions of the slit-truss model. Here we fix the parameters $l_0=1$, $k_s=1$, $l_H=0.125$, $\kappa_\alpha=0.048$, $\kappa_\theta=3\kappa_{\alpha}$.
\textbf{(B)} Buckling for the
closed ($\theta=0$) configurations $C_c$. 
The right buckled branch  includes a range where $\partial{U}/\partial{\theta} > 0$ (green), signaling tensile stresses that would open the slit; these solutions are not admissible.
\textbf{(C)} Imperfect buckling for the 
free ($\theta$ unconstrained) $C_f$
configurations. The snapped branches
include regions where
 $\theta < 0$ (red) which are not admissible. Notice that for finite compression, $\theta=0$ is reached at positive deflection $x$.
\textbf{(D)} 
The combination of the free and constrained configurations captures the bifurcation diagram of slitted beams \textbf{(E)} Zoom in, where the unstable closed (green) and self-overlapping open (red) branches are removed, and the solutions of
the slitted-truss model (blue) are compared to experimental data (gray) for the beam in Fig.~\ref{fig:slit_trend}A.}
\label{fig:spring_sketch}
\end{figure*}

To gain insight into the underlying mechanisms that governs slit-snapping, we develop a truss model
based on the observations that slit beams exhibit buckling in their closed configuration and snap open when the stresses across the slit become tensile. 
We start from the Bellini truss - a minimal model for buckling - which 
consists of two equal linear springs of length $l_o$, coupled by a torsional spring \cite{belliniConcept1972}. To model a right-slit beam, the central Bellini hinge
is replaced by a triplet of hinges connected 
by two additional bars, yielding
a structure that
breaks left-right symmetry (Fig.~\ref{fig:spring_sketch}A). Its geometry is set by the length of the rigid bars $l_H$, and the opening angle $\theta$ of the center hinge that encompasses both the closed ($\theta=0) $ and open
($\theta>0$) configuration of a slit beam.
The slit-truss system is connected at the bottom and top by two freely rotating joints, and is compressed vertically by a strain $\varepsilon = u_y/(2 l_0)$. The compressed slit-truss structure remains top-down symmetric and is characterized by
its mid-point deflection $x$ and angle $\theta$ (Fig.~\ref{fig:spring_sketch}A; for details, see Supplemental Information).

We model the configurations of the spring system by first solving for the closed  ($\theta=0)$ case, $C_c$, and the free ($\theta$ unconstrained) case, $C_f$, and then applying additional criteria derived from the physics of the slit.
For the closed configuration $C_c$, $\theta$ is fixed at zero, and the slit-truss model features
an ordinary, perfect pitchfork bifurcation which represents the buckling of a closed beam (Fig.~\ref{fig:spring_sketch}B).
However, this solution is only admissible  when
compressive torques act on the slit; tensile torques would lead to $\theta>0$. Hence, we require for $C_c$ that the non-holonomic constraint  $\partial E/\partial \theta \le 0$ is satisfied, where $E$ is the potential energy of the spring system. This eliminates part of the 'rb' branch of $C_c$ (Fig.~\ref{fig:spring_sketch}B). 
In the free configuration, $C_f$,
the unconstrained truss model, due to the broken left-right symmetry, features an imperfect pitchfork bifurcation,
producing a 'lb' and a pair of stable and unstable 'rs' solutions (Fig.~\ref{fig:spring_sketch}C). However,
enforcing the non-self-intersecting nature of the slit, these solutions are only admissible when 
$\theta\ge 0$, which eliminates 
parts of the 'lb' and the unstable 'rs' branch of the constrained system (Fig.~\ref{fig:spring_sketch}D).
Combining the admissible solutions produces the bifurcation diagram of slit-truss model, which closely matches the experimental observed behavior of slit beams (Fig.~\ref{fig:spring_sketch}D,E).

The solutions for the closed and free configurations and the admissibility conditions 
clarify the nature of the opening and closing transition of slit beams.
The opening transition occurs at the intersection of
the free and closed solutions, where $\theta$, $\partial E/\partial x$ and $\partial E/\partial \theta$ all become zero. For larger strains, the closed solution turns unstable,  and the configuration discontinuously snaps to the 'rs' branch. Similarly, lowering the strain, the 'rs' branch disappears in a simple saddle-node bifurcation, and the system discontinuously snaps closed.

The slit-truss model futhermore rationalizes the dependence of the opening and closing strains on the beam aspect ratio $t$ and slit size $s$. 
We note that the torsional constants of $R_1$ and $R_2$ can be determined by requiring that they correctly capture the onset of buckling in the $C_c$ configurations, which implies that they scale as $t^2$ \cite{LennardBumping2024}. In addition, $\kappa_{\theta}$ decrease for increasing $s$. 
In our model, the closing strain  is 
controlled by the fold bifurcation of the 
free solution, and we find that, consistent with our experimental and numerical data,  
$\varepsilon_c^{model}$  decreases with
decreasing $k_\theta$, allowing to
establish a relation between $k_\theta$ and $s$ that 
captures the three scenarios shown in Fig.~\ref{fig:slit_trend} (see Supplemental Material). Similarly, as the 
opening strain is determined by the
onset of negative torques in the 
constraint solution, it is independent of $\kappa_\theta$ and thus $s$, consistent with our data. Hence, our model faithfully captures the phenomenology and leading parameter dependencies of slit-snapping.

\subsubsection*{{Beams with dual slits}}

\begin{figure*}
\centering
\includegraphics[width = 6 in]{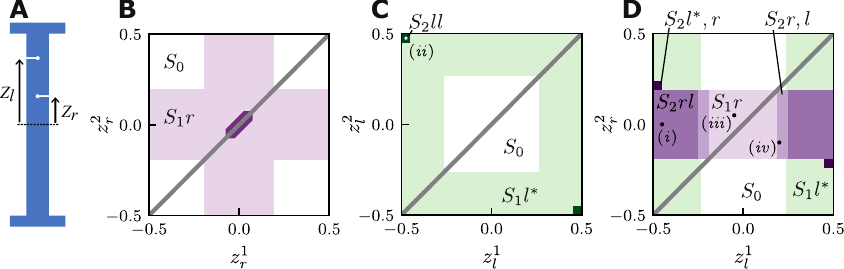}
\caption{
\textbf{Enhanced snapping of dual slit beams.} 
(\textbf{A}) Sketch of a beam with two slits, where $z_r$ and $z_l$ denote the slit positions of a right and left slit respectively. 
(\textbf{B-D}) Snapping responses for beams with two right slits ({\bf B}), a left and a right slit ({\bf C}), and two left slits ({\bf D}). 
The labels and colors denote eight qualitatively distinct behaviors. $S_0$: No snapping. $S_1r$: Single slit snapping of a right slit. $S_2rr$: Joint snapping of two right slits. $S_2rl$: Cooperative snapping of a right and left slit. $S_2r,l$: Sequential snapping (first right slit, then left slit). $S_1l^*$: Smooth opening of the left slit. $S_2l^*, r$: Snapping of the right slit triggered by previous smooth opening of the left slit. $S_2ll$: Cooperative snapping of both left slits.}
\label{fig:4.1}
\end{figure*}

The range and diversity of the snapping behavior of beams with a single slit 
is limited. First, snapping is restricted to a single pair of opening/closing transitions occurring for only one 
of the buckled branches. Furthermore, 
the hysteresis range is 
restricted, and
the closing strain can not be lowered below approximately $0.5\varepsilon_b$. Finally, the snapping transition is unidirectional, always increasing the beam’s deflection.

Here we show that all these limitations can be overcome by using beams with multiple slits. We first
explore the snapping of beams with {\em two} slits using
FEM simulations of beams with fixed thickness and  slit depths ($t = 0.125$, $s = 0.75$, $h = 1$~mm). 
As we will show, snapping of one slit drastically alters the curvature of the beam. Depending on the location of the slits, this can lead to 'interactions' between the slits that are either cooperative, where one snapping event promotes the next, or antagonistic,
where the first snapping event suppresses the second.
We show numerically that 
these interactions 
give rise to a wide variety
of slit-snapping scenario's (Fig.~\ref{fig:4.1}B-D). We then experimentally realize dual-slit beams that materialize the most striking exotic snapping behaviors, and finally explore beams with more than two slits.

We parametrize the vertical positions of the right and left slits by the dimensionless variables $z^i_r$ and $z^i_l$, respectively and examine the three possible arrangements of dual-slit beams: right–right, left–right, and left–left. For each, we discuss and classify the distinct scenarios as function of slit positions, and the corresponding critical strains.
Without loss of generality, we focus on the right-buckled branch; the left-buckled case follows by symmetry.
Beams with two right slits exhibit
three distinct regimes, that we label
$S_0, S_1r$ and $S_2rr$ (Fig.~\ref{fig:4.1}B). 
Here, numerical indices
indicate the number of slits that open and  'l' and 'r' denote which slits participate.
$S_0$: No snapping occurs when both slits lie outside the center region, i.e., when $|z_r^i| > 0.19$, analogous to  the condition for a beam with a single slit (See SI). 
$S_1r$: Ordinary slit-snapping is observed whenever at least one slit is within the center region, i.e., 
has $z_r^1<0.19$. In this regime, the most central slit opens and prevents the other from doing so; the opening and closing strains match those of a single-slit beam.
$S_2rr$: When both $z_r^i \approx 0$, 
the slits cooperate and snap open and close simultaneously.
The opening strain is identical to that of a single slit beam; the closing strain is slightly lower than that of a single slit beam. Hence, the rightward snapping of a beam with a pair of slits at the right side is quite similar to that of a single-slit beam.

In contrast, for left-left slits, right buckling beams 
exhibit three distinct behaviors ($S_0, S_1l^*$ and $S_2ll$). 
(Fig.~\ref{fig:4.1}C). 
$S_0$: When the cuts are in the center region
($|z_l^i|<0.265$), the slits remain closed in right-buckling beams. 
$S_1l^*$: 
When one of the slits 
is located in the boundary region where the curvature is negative ($|z_l^i|<0.265$), it opens smoothly open but does not cause snapping. 
$S_2ll$: Surprisingly, there is a tiny parameter regime where, when both slits are close to the top and bottom boundaries of the beam, we observe right snapping with an opening strain $\varepsilon_o$ lower than for a single slit at $z_r=0$ (Fig.~\ref{fig:4.1}F). Apart from this exotic behavior, beams with left/left slits do not snap to the right.

Finally, for beams with left-right slits,  we observe a wide variety of novel snapping behaviors that we label
$S_2r,l, S_2rl, S_2l^*,r$ and $S_1l^*$; here, 
a comma denotes sequential behavior and stars indicate smooth opening of slits (Fig.~\ref{fig:4.1}D).
We first focus on 
$|z_r| < 0.19$ where three regimes are observed as a function of $z_l$. 
$S_1r$: When $|z_l|$ is small, we observe ordinary  slit snapping, controlled by $z_r$ --- the left slit remains closed and the critical strains are as for a single slit beam. 
$S_2r,l$: 
For intermediate $|z_l|$, we observe two-step snapping: after the right slit snaps open, increasing the compression leads to subsequent snapping of the left slit. Lowering the compression, both slits close sequentially, first the left, then the right. The right slit is the first to open and the last to close, and its critical strains are as for a beam with a single slit. 
$S_2rl$: 
For extreme values of $|z_l|>0.23$, both slits open and close simultaneously and cooperatively. When the right slit opens, the resulting change in beam curvature immediately triggers the opening of the left slit; 
the opening strain is as for a single slit beam. However, the closing strain is significantly lowered: the opening of the  
left slit prevents the right slit to close.

For large $|z_r|>0.19$, three additional regimes are found. 
$S_0$: 
For small $|z_l|<0.265$, we observe no snapping. 
$S_1l^*$: For larger  
$|z_l|$, the left slit opens up continuously, and the right slit remains closed in most of parameter space. 
$S_2l^*r$: There is a small region of parameter space where the smooth opening of the left slit eventually triggers the snapping of the right slit. In this regime, both the opening and closing strains are modified from their single-slit values.

We cluster the snapping behavior of beams with two slits in 
three groups. First,  in significant swaths of parameter space, only one slit participates in the snapping behavior as the other remains closed  ($S_1r$).
Second, in the regimes
$S_2rr, S_2rl$ and $S_2r,l$
the first snapping event is triggered by the instability of a single slit, and $\varepsilon_o$ is as for single-slit beams. However, this initial event then triggers a subsequent snapping of the other slit, allowing to either lower $\varepsilon_c$ or to
realize sequential slit-snapping. 
Third, in extreme cases ($S_2l^*,r$ and $S_2ll$), snapping is a cooperative effect and occurs for slit positions  where a single slit would not produce snapping.

Taken together, these results suggest that multiple slits allow to significantly extend the range of slit-snapping behaviors.

\subsubsection*{Experimental Realization of Exotic Snapping}

\begin{figure*}
\centering
\includegraphics[width = 6 in]{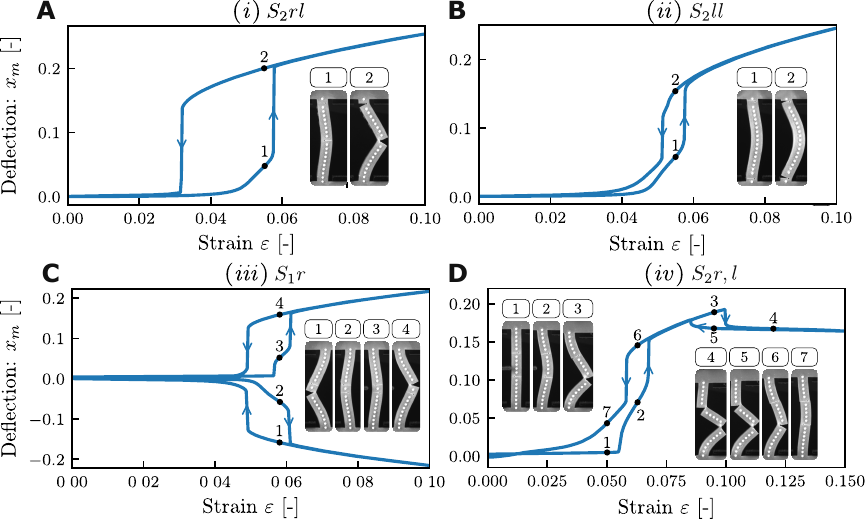}
\caption{
\textbf{
Experimental realization of four exotic slit snapping scenarios in dual slit beams.} Beam deflection vs. strain with snapshots of distinct beam configurations.
(\textbf{A}) 
Beam with large hysteresis range (regime $S_2rl$, $z_l=0.45$, $z_r=0$).
(\textbf{B}) Beam with two left slits that snaps to the right (regime $S_2ll$, $z_l=-0.475,0.475$). Parameter values are also indicated in Fig.~\ref{fig:4.1} as {\em (i)-(iv)}. 
(\textbf{C}) 
Beam that snaps both to the left and to the right
(regime $S_1r$, $z_l=-0.05$, $z_r=0.05$).
(\textbf{D}) Beam with sequential snapping and unsnapping (regime $S_2r,l$, $z_l=0.2$, $z_r=-0.1$).
}  
\label{fig:4.2}
\end{figure*}

To demonstrate 
the potential of dual-slit beams experimentally, 
we realized the four most striking
slit-snapping scenarios. To do so, we select four representative values for the slit positions  based on the numerical simulations, fabricate the corresponding 
samples,  and characterize their behavior (Fig.~\ref{fig:4.2}).

{\em (i)} First, we realize a beam with a large hysteresis loop in the $S2_{rl}$ regime, where a left slit near the beam’s end ($z_l = 0.45$) and a centered right slit ($z_r = 0$) act cooperatively
to produce this response (Fig.~\ref{fig:4.2}A and movie S3).
In essence, the increase in curvature induced by the snapping of the right slit at $\epsilon_o$ results in the simultaneous opening of the left slit, which in turn lowers the curvature of the top half of the beam (Fig.~\ref{fig:4.2}A).
This lowered curvature suppreses the closing of the beam and strongly lowers   
the critical strain $\varepsilon_c$. Hence, dual-slit beams can significantly enlarge the hysteresis range of slit beams. 

{\em (ii)} Second, we fabricate a sample in the $S2ll$ regime, where two left slits at the extremes of a beam, $(z_l^1, z_l^2)=(0.475,-0.475)$, can cause rightward slit-snapping (Fig.~\ref{fig:4.2}B and movie S4). 
While the hysteresis loop of this sample resembles that of a single-slit beam, its 
post-snapping configuration
is markedly different, lacking inflection points and resembling
a buckled beam with pinned boundaries.
Moreover, the snapping here requires both slits to be present, in contrast to case {\em (i)} where the snapping is triggered by the right slit independently of the presence of the left slit. Hence, scenario {\em (iv)} highlights that cooperation between slits can give rise to qualitatively distinct  instabilities.

{\em (iii)} Third, we realize a quadstable 
beam, by using 
a left and a right slit to allow
snapping of both the right and left buckled branches   (Fig.~\ref{fig:4.2}C and movies S5 and S6). To select the slit positions, we note that point reflection symmetry of the design space about the origin implies that similar single-slit snapping responses can be selected for both the left and right buckling branches by taking 
$z_r^1 = -z_l^2$; here we take 
a design inside the $S1$ region ($(z_r^1, z_l^2)=(0.05,-0.05)$). 
When the beam buckles right (left), only the right (left) slit snaps, and the left (right) slit remains closed (Fig.~\ref{fig:4.2}C).
Hence, this dual-slit beam can snap both to the left and to the right, mimicking the 
behavior of a single slit beam with either $z_r=0.05$
or $z_l=-0.05$. We note that the slits do not cooperate, and this would  allow to tune the snapping of their left and right buckled branch independently by, e.g., varying the corresponding slit depth.

{\em (iv)} Finally, we realize sequential snapping with a beam  in the $S2r,l$ regime ($(z_r^1, z_l^2)=(-0.1,0.2)$;  Fig.~\ref{fig:4.2}D and movie S7).
Under increasing strain, the beam initially exhibits regular snapping of the right slit while the left slit remains closed. As the strain increases further, the left slit snaps at a second snapping strain, leading to an S-shaped configuration in which both slits are open (Fig.~\ref{fig:4.2}D).
Both snapping transitions are hysteretic,
and as the strain is decreased, first the left, then the right slit closes. We note that in our experiments
the second slit closing 
leads to a
small transversal
mismatch between the top and bottom parts of the slit, making the smooth return to the unbuckled configuration hysteretic, as the mismatch is gradually disappears as the strain is removed (Fig.~\ref{fig:4.2}D).
Hence, dual slits allow sequential snapping in a single beam.

Together, these four scenarios demonstrate the effectiveness of introducing an additional slit as a strategy to experimentally enhance and tune the snapping and response of slit beams. Moreover, the geometric understanding of slit–slit interactions provides a foundation for the rational design of targeted responses in beams with more than two slits, where exhaustive combinatorial exploration becomes impractical.

\subsubsection*{Beams with Many Slits and Extreme Pathways}

\begin{figure*}
\centering
\includegraphics{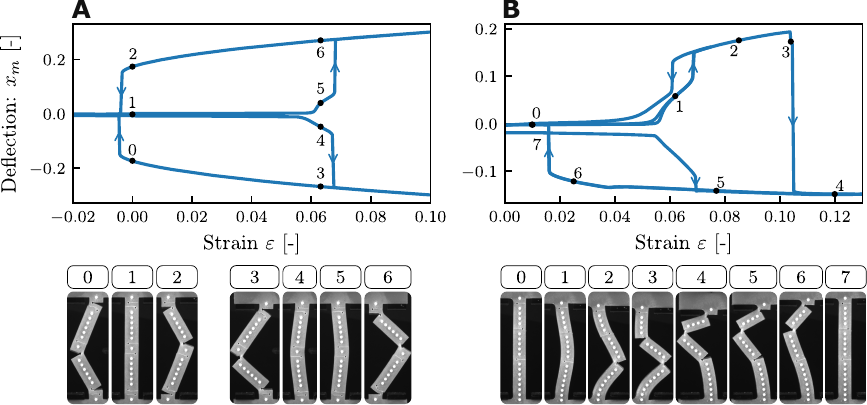}
\caption{\textbf{Complex snapping with multiple slits.}
(\textbf{A}) A beam with six deep slits ($t=0.1375,  s=0.8, z_l=-0.45,-0.05,0.4, z_r=-0.4,0.05,0.45$) shows tristability at zero strain. (\textbf{B}) A beam with three slits ($t=0.125,  s=0.75, z_l=0.20, z_r=-0.1,0.45$) snaps from the right buckling branch to the left during compression.}
\label{fig:final2}
\end{figure*}

To demonstrate the design potential unlocked by using multiple slits, we experimentally realize two beams with many slits that exhibit targeted exotic behaviors. First, we realize a beam that is tristable at {\em zero} strain using six slits, and second we realize a beam with  a snapping transition from the right to the left buckling branch using three slits
(Fig.~\ref{fig:final2}).

To realize a slit-beam that is 
tristable at zero compressive strain, we combine the design ideas for realizing enhanced hysteresis (Fig.~\ref{fig:4.2}A)
and for snapping in both branches 
(Fig.~\ref{fig:4.2}C).
We first focus on the enhanced hysteresis, and aim to create a beam that is bistable at zero strain, i.e., pushing $\varepsilon_c$ to negative values (Fig.~\ref{fig:final2}A and movies S8 and S9).
To do so, we increase the slit depths to $s = 0.8$, and investigate a beam with three slits, at $z_r^i = (0.05)$ and
$z_l^i = (-0.45, 0.4)$. Here the snapping is initiated by the right slit, which then triggers the opening of  both left slits simultaneously, leading to 
a post-snap configuration where the curvature is strongly concentrated near the slits so that the beam segments are nearly straight; this can be seen as a more extreme version of enlarged hysteresis range for the dual-slit beam (Fig.~\ref{fig:4.2}A). 
Here, this straightening of the beam segments lowers the collective closing of the three slits to negative strain values ($\varepsilon_c < 0$) and makes the right branch bistable at zero strain (Fig.~\ref{fig:final2}A). To achieve tristability at zero strain, we
induce three additional slits near $z_l^i = (-0.05)$ and $z_r^i = (0.45, -0.4)$, that satisfy 
the symmetry  $z_l \rightarrow -z_r$ and $z_r \rightarrow -z_l$ as described above.
Notice that this requires to slightly break the top-down symmetry of each set of three slits to avoid pairs of left and right slits at the same vertical position. 
While the slits in either of these two sets act cooperatively, the vertical alignment of the two sets induces strong anti-cooperative interactions between opposing pairs and only one set is opened at each of the left and right snapped branches.
Note that this approach also readily enables the creation of tristable beams with asymmetric branches by modifying slits from one of the sets, thereby allowing independent programming of each branch.

We finally realize a beam with three slits that snaps from the right to the left buckling branch during compression. To program this response, we enhance the sequential snapping behavior observed in {\em(iii)} with an additional slit that allows the beam to overcome
the energy barrier that separates the
left and right
branches (Fig.~\ref{fig:final2}B and movie S10). The beam features three slits of depth $s=0.75$, one cut from the left and two from the right at $z_l^i = (0.2)$, $z_r^i = (0.45, -0.1)$. To systematically observe right buckling, we weakly break the left-right symmetry of the beam by introducing two small defects at the ends of the beams, which allows us to avoid the use of a static indenter. During compression, the beam first buckles right,  after which the first slit-snapping instability that opens the bottom slit
is triggered. The two other slits remain closed and lowering the strain leads to a unsnapping transition analogous to a single slit beam (Fig.~\ref{fig:final2}B). 

However, further compression increases the curvature near the middle left slit, eventually producing a second slit-snap analogous to the phenomenon shown in {\em(iii)}. This second instability changes the sign of the curvature near the top of the beam and 
triggers the opening of the top slit, which causes the beam to snap from the right to the left branch via an intermediate S-shaped configuration, reminiscent of the second buckling mode \cite{Colin2025singletoggleron}. 
The three slits are open in the final left branch configuration. When the strain is quasistatically removed, first the bottom slit closes smoothly, resulting in a kink in the stability diagram, and then the two remaining slits collectively close at a lower strain, returning to the unbuckled configuration. From this state, the snapping sequence can be re-initialized by increasing the strain. 

These results highlight the broader potential of slit beams—with many slits—as a platform for encoding tailored multistability and directional instabilities.


\subsection*{{DISCUSSION}}

\subsubsection*{}

Endowing flexible structures with partial cuts that open and close as a result of global compressive stresses allow for a powerful collaboration between
geometric and contact nonlinearities. For beams, we have uncovered the principle mechanisms that govern slit-enhanced snapping, and demonstrated that it yields a combination of continuous buckling and discontinuous snapping under compression, large and tunable hysteresis cycles, multistability and sequential behavior - all in a single slit beam. Interesting extensions include the use of slits for other geometries, and the use of multiple interacting slit-snapping elements.

We close by briefly discussing three general areas of applications. First, snapping beams have been used to
create materials that store and dissipate energy \cite{Damiano2015SnappingUnderTension, shanMultistable2015, yangMultistable2019}
and store information \cite{LennardCounter2023, Reis2021MagnetoMechanical, Colin2025togglerons, Jerry2024}.
As the number of stable states directly links with the energy and memory capacity, slits can be used to increase these capacities without changing the number of elements. Moreover, since energies and memories are often stored in buckled structures states~\cite{LennardCounter2023, Colin2025togglerons, Colin2025pathdependency, Katia2016Propagation, shanMultistable2015}, slit beams offer a simple way to tune or qualitatively modify their storage capacity \cite{ChloeLatch2025, Chloe2021}.

Second, bistable elements have recently been used for information processing, with sequential driving encoding input and transitions between multistable states providing output — a concept that relies critically on hysteretic elements or hysterons \cite{Nathan2020GlobalMemoryFromLocalHysteresis, Hadrien2021, Dor2022Crumpled, LennardCounter2023, Jerry2024, Colin2025togglerons, Joey2025, DorTriangularHub, ChloeLatch2025, Chloe2021, NathanJoey2025Review, Hecke2021Profussion}
Mechanical implementations often rely on buckling under axial compression to store the information and lateral forcing to change it; however, slit beams allow to control buckling and snapping under the same type of forcing, opening up the possibility for simpler and more powerful driving protocols. Moreover, as slit-snapping concentrates  deformations perpendicular to the driving, it facilitates 
the propagation of signals by interactions between neighboring elements \cite{Katia2016Propagation,LennardCounter2023}. Finally, 
as the interactions between multiple slits can both be cooperative or anti-cooperative, similar to the interactions needed to access complex sequences able to store memory and process information, slit-beams could form multi-bit building blocks for more advanced computing  \cite{DorTriangularHub, Colin2025pathdependency, Margot2024transitiongraphs, ChloeLatch2025, Hadrien2021, Yasuda2021MechanicalComputing}.


Third, physical intelligence in, e.g., 
soft robots or advanced metamaterials, often relies on instabilities and multistability, where switching between states can be triggered by interactions with the environment \cite{KatiaRobot2024, BasComoretto2025embodyingmechanofluidicmemorysoft}
We expect that our simple strategy with partial cuts will find applications in making soft robots more adaptive, faster, smarter and more easy to design. 

\section*{Acknowledgments}
The authors acknowledge Dion Ursem and Jeroen Mesman for technical support and thank Colin Meulblok, Paul Ducarme and Sourav Roy for insightfull discussions. B.D.F. and M.v.H. acknowledge funding from the European Research Council Grant ERC-101019474.


\section*{Materials and Methods}

\subsubsection*{Manufacturing the beams}

To manufacture slitted beams, we 3D print open-face molds to the beam's desired dimensions, including a pin to create the hole. We then cast the molds using Smooth-On Mold Star 30 VPS rubber. A roughly incompressible rubber with a shore hardness of 30A, Youngs Modulus of Young's Modulus E $\approx 738~\text{kPa}$ and density $\rho=1120kg/m^3$ \cite{zehnderMetasilicone2017}. We allow the silicone to cure for at least one day before demolding. After demolding, we cut the slit using a custom device that features a scalpel attached to a linear slider rail.

\subsubsection*{Uniaxial compression set-up}
We perform experiments in a custom build set-up consisting of two parallel horizontal plates with controllable relative vertical distance in which we mount the beam, and a horizontal indenter (Fig.~S1). 
We mount the beams using a pair of 3D printed, horizontally aligned clamps that tightly fit the support of the beams (Fig.~S1). 
We apply talc powder between the top and bottom parts of the slit of the beam, as well as between the indenter and the beam, to reduce the stickiness at these contact points.

The bottom plate position is fixed, while the top plate can be moved in the $y$-direction with an accuracy of $\pm 0.01 \, \text{mm}$ using an stepper motor connected to a computer. The plates are mounted on rails that ensure that the angle between plates is smaller than $0.6 \, \text{mm/m}$. We calibrate the vertical distance between the plates using a copper beam of length $140 \pm 0.03 \,  \text{mm}$; a dial indicator with sensitivity $\pm 0.01 \, \text{mm}$ is used to check that the plates remain well calibrated during 
experiments.

\subsubsection*{Clamping}
We employ clamped-clamped boundary conditions on the beams by casting them with two support blocks at each end that fit a pair of clamps attached to the plates. The clamps are aligned in the $x-z$ plane by bringing them together before screwing them tightly to the plates. Calibration in $y$ is achieved by pushing them towards the back of the clamps (Fig.~S1).

\subsubsection*{Symmetry breaking}
To control the initial buckling direction, we use a 3D printed indenter capped with a semi-circular tip of radius 5 mm to set a finite midpoint deflection and explicitly break left-right symmetry. The indenter is attached to a manual linear stage that controls its $x$-position with an accuracy of $\pm 0.01 \, \text{mm}$ (Fig.~S1). The indenter pushes the beam laterally at height $L/2$, and we manually adjust its $x$-position to set a minimum midpoint deflection $x_{in}$. Experiments show that once the beam loses contact with the indenter, the effect of $x_{in}$ on the beam configuration is minimal (Fig.~S2); in practice we use $x_{in}$ of order 1 mm . When measures of both the right and left branch of a beam are performed, we first let the beam buckle spontaneously, then we measure the other branch by using the indenter. This can result in small asymmetries near the buckling strain between the branches (Fig.~\ref{fig:4.2}).

\subsubsection*{Data acquisition}

A grayscale CMOS camera records the experiments at $3 \, \text{Hz}$ with a resolution of resolution of 3088 × 2064 pixels, yielding images with a pixel density of $5~px/mm$. The beams feature tiny circular bumps that are painted white along their center line (Fig.~S1). We track the position of these dots along the beam with accuracy $\pm 0.02\text{mm}$ using a custom script based on the library opencv \cite{opencv_library}. We use the mean $x$-distance between the two central dots and the neutral line to compute the midpoint lateral deflection $x_m$. 

\subsubsection*{Numerical simulations}
We complement the experiments with finite element (FEM) simulations to tackle two limitations of the experiments; material induced hysteresis steaming from material creep (Fig.~\ref{fig:slit_trend}A) and limited reproducibility when re-clamping the specimens.
We perform simulations in the software package Abaqus, using the Dynamic/Explicit solver, 2D geometry with plain stress, a Neo-Hookean material with a Poisson ratio $\nu=0.49$, Young's modulus $E=0.78$ MPa and sufficient damping to avoid oscillations.



\clearpage 

%
\bibliography{SlitBeams} 

\begin{thebibliography}{10}
\providecommand{\url}[1]{\texttt{#1}}
\expandafter\ifx\csname urlstyle\endcsname\relax
  \providecommand{\doi}[1]{doi:\discretionary{}{}{}#1}\else
  \providecommand{\doi}{doi:\discretionary{}{}{}\begingroup \urlstyle{rm}\Url}\fi

\bibitem{bertolidmetareview2017}
K.~Bertoldi, V.~Vitelli, J.~Christensen, M.~van Hecke, Flexible mechanical metamaterials. \emph{Nature Reviews Materials} \textbf{2}~(11), 17066 (2017), \doi{10.1038/natrevmats.2017.66}, \url{https://doi.org/10.1038/natrevmats.2017.66}.

\bibitem{KatiaReis2010NegativePoisson}
K.~Bertoldi, P.~M. Reis, S.~Willshaw, T.~Mullin, Negative Poisson's Ratio Behavior Induced by an Elastic Instability. \emph{Advanced Materials} \textbf{22}~(3), 361--366 (2010), \doi{https://doi.org/10.1002/adma.200901956}, \url{https://advanced.onlinelibrary.wiley.com/doi/abs/10.1002/adma.200901956}.

\bibitem{Corentin2015DiscontinuousBuckling}
C.~Coulais, J.~T.~B. Overvelde, L.~A. Lubbers, K.~Bertoldi, M.~van Hecke, Discontinuous Buckling of Wide Beams and Metabeams. \emph{Phys. Rev. Lett.} \textbf{115}, 044301 (2015), \doi{10.1103/PhysRevLett.115.044301}, \url{https://link.aps.org/doi/10.1103/PhysRevLett.115.044301}.

\bibitem{sachseSnapping2020}
R.~Sachse, \emph{et~al.}, Snapping Mechanics of the {{Venus}} Flytrap ( {{{\emph{Dionaea}}}}{\emph{ Muscipula}} ). \emph{Proceedings of the National Academy of Sciences} \textbf{117}~(27), 16035--16042 (2020), \doi{10.1073/pnas.2002707117}, \url{https://pnas.org/doi/full/10.1073/pnas.2002707117}.

\bibitem{holmesSnapping2007}
D.~P. Holmes, A.~J. Crosby, Snapping {{Surfaces}}. \emph{Advanced Materials} \textbf{19}~(21), 3589--3593 (2007), \doi{10.1002/adma.200700584}, \url{https://onlinelibrary.wiley.com/doi/10.1002/adma.200700584}.

\bibitem{bennet-clarkEnergetics1975}
H.~C. {Bennet-Clark}, The Energetics of the Jump of the Locust {{{\emph{Schistocerca}}}}{\emph{ Gregaria}}. \emph{Journal of Experimental Biology} \textbf{63}~(1), 53--83 (1975), \doi{10.1242/jeb.63.1.53}, \url{https://journals.biologists.com/jeb/article/63/1/53/22130/The-energetics-of-the-jump-of-the-locust}.

\bibitem{queathemOntogeny1991}
E.~Queathem, The Ontogeny of Grasshopper Jumping Performance. \emph{Journal of Insect Physiology} \textbf{37}~(2), 129--138 (1991), \doi{10.1016/0022-1910(91)90098-K}, \url{https://linkinghub.elsevier.com/retrieve/pii/002219109190098K}.

\bibitem{jinUltrafast2023}
L.~Jin, \emph{et~al.}, Ultrafast, {{Programmable}}, and {{Electronics}}-{{Free Soft Robots Enabled}} by {{Snapping Metacaps}}. \emph{Advanced Intelligent Systems} \textbf{5}~(6), 2300039 (2023), \doi{10.1002/aisy.202300039}, \url{https://onlinelibrary.wiley.com/doi/10.1002/aisy.202300039}.

\bibitem{overveldeAmplifying2015}
J.~T.~B. Overvelde, T.~Kloek, J.~J.~A. D'haen, K.~Bertoldi, Amplifying the Response of Soft Actuators by Harnessing Snap-through Instabilities. \emph{Proceedings of the National Academy of Sciences} \textbf{112}~(35), 10863--10868 (2015), \doi{10.1073/pnas.1504947112}, \url{https://pnas.org/doi/full/10.1073/pnas.1504947112}.

\bibitem{yangPhasetransforming2016}
D.~Yang, \emph{et~al.}, Phase-Transforming and Switchable Metamaterials. \emph{Extreme Mechanics Letters} \textbf{6}, 1--9 (2016), \doi{10.1016/j.eml.2015.11.004}, \url{https://linkinghub.elsevier.com/retrieve/pii/S2352431615300092}.

\bibitem{yangMultistable2019}
H.~Yang, L.~Ma, Multi-Stable Mechanical Metamaterials by Elastic Buckling Instability. \emph{Journal of Materials Science} \textbf{54}~(4), 3509--3526 (2019), \doi{10.1007/s10853-018-3065-y}, \url{http://link.springer.com/10.1007/s10853-018-3065-y}.

\bibitem{dingSequential2022}
J.~Ding, M.~Van~Hecke, Sequential Snapping and Pathways in a Mechanical Metamaterial. \emph{The Journal of Chemical Physics} \textbf{156}~(20), 204902 (2022), \doi{10.1063/5.0087863}, \url{https://pubs.aip.org/jcp/article/156/20/204902/2841495/Sequential-snapping-and-pathways-in-a-mechanical}.

\bibitem{rafsanjaniBucklinginduced2017}
A.~Rafsanjani, K.~Bertoldi, Buckling-{{Induced Kirigami}}. \emph{Physical Review Letters} \textbf{118}~(8), 084301 (2017), \doi{10.1103/PhysRevLett.118.084301}, \url{https://link.aps.org/doi/10.1103/PhysRevLett.118.084301}.

\bibitem{rafsanjaniSnapping2015}
A.~Rafsanjani, A.~Akbarzadeh, D.~Pasini, Snapping {{Mechanical Metamaterials}} under {{Tension}}. \emph{Advanced Materials} \textbf{27}~(39), 5931--5935 (2015), \doi{10.1002/adma.201502809}, \url{https://onlinelibrary.wiley.com/doi/10.1002/adma.201502809}.

\bibitem{shanMultistable2015}
S.~Shan, \emph{et~al.}, Multistable {{Architected Materials}} for {{Trapping Elastic Strain Energy}}. \emph{Advanced Materials} \textbf{27}~(29), 4296--4301 (2015), \doi{10.1002/adma.201501708}, \url{https://onlinelibrary.wiley.com/doi/10.1002/adma.201501708}.

\bibitem{Reis2021MagnetoMechanical}
T.~Chen, M.~Pauly, P.~M. Reis, A reprogrammable mechanical metamaterial with stable memory. \emph{Nature} \textbf{589}~(7842), 386--390 (2021), \doi{10.1038/s41586-020-03123-5}, \url{https://doi.org/10.1038/s41586-020-03123-5}.

\bibitem{Katia2022}
D.~Melancon, A.~E. Forte, L.~M. Kamp, B.~Gorissen, K.~Bertoldi, Inflatable Origami: Multimodal Deformation via Multistability. \emph{Advanced Functional Materials} \textbf{32}~(35), 2201891 (2022), \doi{https://doi.org/10.1002/adfm.202201891}, \url{https://advanced.onlinelibrary.wiley.com/doi/abs/10.1002/adfm.202201891}.

\bibitem{meeussen}
A.~S. Meeussen, M.~van Hecke, Multistable sheets with rewritable patterns for switchable shape-morphing. \emph{Nature} \textbf{621}~(7979), 516--520 (2023), \doi{10.1038/s41586-023-06353-5}, \url{https://doi.org/10.1038/s41586-023-06353-5}.

\bibitem{LennardCounter2023}
L.~J. Kwakernaak, M.~van Hecke, Counting and Sequential Information Processing in Mechanical Metamaterials. \emph{Phys. Rev. Lett.} \textbf{130}, 268204 (2023), \doi{10.1103/PhysRevLett.130.268204}, \url{https://link.aps.org/doi/10.1103/PhysRevLett.130.268204}.

\bibitem{Jerry2024}
J.~Liu, \emph{et~al.}, Controlled pathways and sequential information processing in serially coupled mechanical hysterons. \emph{Proceedings of the National Academy of Sciences} \textbf{121}~(22), e2308414121 (2024), \doi{10.1073/pnas.2308414121}, \url{https://www.pnas.org/doi/abs/10.1073/pnas.2308414121}.

\bibitem{LennardBumping2024}
L.~J. Kwakernaak, A.~Guerra, D.~P. Holmes, M.~{van Hecke}, The collective snapping of a pair of bumping buckled beams. \emph{Extreme Mechanics Letters} \textbf{69}, 102160 (2024), \doi{https://doi.org/10.1016/j.eml.2024.102160}, \url{https://www.sciencedirect.com/science/article/pii/S2352431624000403}.

\bibitem{HolmesOyster2023}
A.~Guerra, A.~C. Slim, D.~P. Holmes, O.~Kodio, Self-Ordering of Buckling, Bending, and Bumping Beams. \emph{Phys. Rev. Lett.} \textbf{130}, 148201 (2023), \doi{10.1103/PhysRevLett.130.148201}, \url{https://link.aps.org/doi/10.1103/PhysRevLett.130.148201}.

\bibitem{KatiaRobot2024}
L.~M. Kamp, \emph{et~al.}, Reprogrammable sequencing for physically intelligent under-actuated robots (2024), \url{https://arxiv.org/abs/2409.03737}.

\bibitem{Colin2025pathdependency}
C.~M. Meulblok, A.~Singh, M.~Labousse, M.~van Hecke, Path-dependency and emergent computing under vectorial driving (2025), \url{https://arxiv.org/abs/2503.07764}.

\bibitem{Yasuda2021MechanicalComputing}
H.~Yasuda, \emph{et~al.}, Mechanical computing. \emph{Nature} \textbf{598}~(7879), 39--48 (2021), \doi{10.1038/s41586-021-03623-y}, \url{https://doi.org/10.1038/s41586-021-03623-y}.

\bibitem{belliniConcept1972}
P.~X. Bellini, The Concept of Snap-Buckling Illustrated by a Simple Model. \emph{International Journal of Non-Linear Mechanics} \textbf{7}~(6), 643--650 (1972), \doi{10.1016/0020-7462(72)90004-2}, \url{https://linkinghub.elsevier.com/retrieve/pii/0020746272900042}.

\bibitem{Colin2025singletoggleron}
C.~M. Meulblok, H.~Bense, M.~Caelen, M.~van Hecke, Accelerated snapping of slender beams under lateral forcing (2025), \url{https://arxiv.org/abs/2505.10091}.

\bibitem{Damiano2015SnappingUnderTension}
A.~Rafsanjani, A.~Akbarzadeh, D.~Pasini, Snapping Mechanical Metamaterials under Tension. \emph{Advanced Materials} \textbf{27}~(39), 5931--5935 (2015), \doi{https://doi.org/10.1002/adma.201502809}, \url{https://advanced.onlinelibrary.wiley.com/doi/abs/10.1002/adma.201502809}.

\bibitem{Colin2025togglerons}
C.~M. Meulblok, M.~van Hecke, Transients and multiperiodic responses: a hierarchy of material bits (2025), \url{https://arxiv.org/abs/2505.09517}.

\bibitem{Katia2016Propagation}
J.~R. Raney, \emph{et~al.}, Stable propagation of mechanical signals in soft media using stored elastic energy. \emph{Proceedings of the National Academy of Sciences} \textbf{113}~(35), 9722--9727 (2016), \doi{10.1073/pnas.1604838113}, \url{https://www.pnas.org/doi/abs/10.1073/pnas.1604838113}.

\bibitem{ChloeLatch2025}
C.~W. Lindeman, T.~R. Jalowiec, N.~C. Keim, Generalizing multiple memories from a single drive: The hysteron latch. \emph{Science Advances} \textbf{11}~(5), eadr5933 (2025), \doi{10.1126/sciadv.adr5933}, \url{https://www.science.org/doi/abs/10.1126/sciadv.adr5933}.

\bibitem{Chloe2021}
C.~W. Lindeman, S.~R. Nagel, Multiple memory formation in glassy landscapes. \emph{Science Advances} \textbf{7}~(33), eabg7133 (2021), \doi{10.1126/sciadv.abg7133}, \url{https://www.science.org/doi/abs/10.1126/sciadv.abg7133}.

\bibitem{Nathan2020GlobalMemoryFromLocalHysteresis}
N.~C. Keim, J.~Hass, B.~Kroger, D.~Wieker, Global memory from local hysteresis in an amorphous solid. \emph{Phys. Rev. Res.} \textbf{2}, 012004 (2020), \doi{10.1103/PhysRevResearch.2.012004}, \url{https://link.aps.org/doi/10.1103/PhysRevResearch.2.012004}.

\bibitem{Hadrien2021}
H.~Bense, M.~van Hecke, Complex pathways and memory in compressed corrugated sheets. \emph{Proceedings of the National Academy of Sciences} \textbf{118}~(50), e2111436118 (2021), \doi{10.1073/pnas.2111436118}, \url{https://www.pnas.org/doi/abs/10.1073/pnas.2111436118}.

\bibitem{Dor2022Crumpled}
D.~Shohat, D.~Hexner, Y.~Lahini, Memory from coupled instabilities in unfolded crumpled sheets. \emph{Proceedings of the National Academy of Sciences} \textbf{119}~(28), e2200028119 (2022), \doi{10.1073/pnas.2200028119}, \url{https://www.pnas.org/doi/abs/10.1073/pnas.2200028119}.

\bibitem{Joey2025}
J.~D. Paulsen, Mechanical hysterons with tunable interactions of general sign (2025), \url{https://arxiv.org/abs/2409.07726}.

\bibitem{DorTriangularHub}
D.~Shohat, M.~van Hecke, Geometric Control and Memory in Networks of Hysteretic Elements. \emph{Phys. Rev. Lett.} \textbf{134}, 188201 (2025), \doi{10.1103/PhysRevLett.134.188201}, \url{https://link.aps.org/doi/10.1103/PhysRevLett.134.188201}.

\bibitem{NathanJoey2025Review}
J.~D. Paulsen, N.~C. Keim, Mechanical Memories in Solids, from Disorder to Design. \emph{Annual Review of Condensed Matter Physics} \textbf{16}~(Volume 16, 2025), 61--81 (2025), \doi{https://doi.org/10.1146/annurev-conmatphys-032822-035544}, \url{https://www.annualreviews.org/content/journals/10.1146/annurev-conmatphys-032822-035544}.

\bibitem{Hecke2021Profussion}
M.~van Hecke, Profusion of transition pathways for interacting hysterons. \emph{Phys. Rev. E} \textbf{104}, 054608 (2021), \doi{10.1103/PhysRevE.104.054608}, \url{https://link.aps.org/doi/10.1103/PhysRevE.104.054608}.

\bibitem{Margot2024transitiongraphs}
M.~H. Teunisse, M.~van Hecke, Transition Graphs of Interacting Hysterons: Structure, Design, Organization and Statistics (2024), \url{https://arxiv.org/abs/2404.11344}.

\bibitem{BasComoretto2025embodyingmechanofluidicmemorysoft}
A.~Comoretto, T.~Mandke, J.~T.~B. Overvelde, Embodying mechano-fluidic memory in soft machines to program behaviors upon interactions (2025), \url{https://arxiv.org/abs/2502.19192}.

\bibitem{zehnderMetasilicone2017}
J.~Zehnder, E.~Knoop, M.~B{\"a}cher, B.~Thomaszewski, Metasilicone: Design and Fabrication of Composite Silicone with Desired Mechanical Properties. \emph{ACM Transactions on Graphics} \textbf{36}~(6), 1--13 (2017), \doi{10.1145/3130800.3130881}, \url{https://dl.acm.org/doi/10.1145/3130800.3130881}.

\bibitem{opencv_library}
G.~Bradski, The {{OpenCV}} Library. \emph{Dr. Dobb's Journal of Software Tools}  (2000).

\end{thebibliography}
\bibliographystyle{sciencemag}

%
%
%
%
%
%

\end{document}